\documentclass{pasj00}

\def\commenta{$^*$}
\def\commentb{$^\dagger$}
\def\commentc{$^\ddagger$}
\def\commentd{$^\S$}
\def\commente{$^\|$}

\def\submitted{submitted}
\def\inpress{in press}
\def\astroph#1{ (astro-ph/#1)}

\DeclareAbbreviation\AAHam{Astron. Abh. Hamburg. Sternw.}
\DeclareAbbreviation\AARv{Astron. Astrophys. Rev.}
\DeclareAbbreviation\an{Astron. Nachr.}
\DeclareAbbreviation\AcA{Acta Astron.}
\DeclareAbbreviation\Afz{Astrofizika}
\DeclareAbbreviation\AnTok{Tokyo Astron. Obs. Annals, Sec. Ser.}
\DeclareAbbreviation\Ap{Astrophysics}
\DeclareAbbreviation\ARep{Astron. Rep.}
\DeclareAbbreviation\ATel{Astronomer's Telegram}
\DeclareAbbreviation\ATsir{Astron. Tsirk.}
\DeclareAbbreviation\AcApS{Acta Astrophys. Sinica}
\DeclareAbbreviation\AstL{Astron. Letters}
\DeclareAbbreviation\BaltA{Baltic Astron.}
\DeclareAbbreviation\BASI{Bull. Astron. Soc. India}
\DeclareAbbreviation\BeSN{Be Star Newsletter}
\DeclareAbbreviation\GCN{GCN}
\DeclareAbbreviation\ibvs{Inf. Bull. Variable Stars}
\DeclareAbbreviation\JAD{J. Astron. Data}
\DeclareAbbreviation\JAVSO{J. American Assoc. Variable Star Obs.}
\DeclareAbbreviation\JBAA{J. British Astron. Assoc.}
\DeclareAbbreviation\LowOB{Lowell Obs. Bull.}
\DeclareAbbreviation\MitVS{Mitteil. Ver\"{a}nderl. Sterne}
\DeclareAbbreviation\MmSAI{Mem. Soc. Astron. Ita.}
\DeclareAbbreviation\Msngr{Messenger}
\DeclareAbbreviation\NewA{New Astron.}
\DeclareAbbreviation\NewAR{New Astron. Rev.}
\DeclareAbbreviation\OAP{Odessa Astron. Publ.}
\DeclareAbbreviation\Obs{Observatory}
\DeclareAbbreviation\PASA{Publ. Astron. Soc. Australia}
\DeclareAbbreviation\PAZh{Pis'ma AZh}
\DeclareAbbreviation\PhR{Phys. Rep.}
\DeclareAbbreviation\PVSS{Publ. Variable Stars Sect. R. Astron. Soc. New Zealand}
\DeclareAbbreviation\PZ{Perem. Zvezdy}
\DeclareAbbreviation\PZP{Perem. Zvezdy Pril.}
\DeclareAbbreviation\QJRAS{QJRAS}
\DeclareAbbreviation\RMxAA{Rev. Mexicana Astron. Astrof.}
\DeclareAbbreviation\RvMA{Reviews of Modern Astron.}
\DeclareAbbreviation\Sci{Science}
\DeclareAbbreviation\SvA{Soviet Astronomy}
\DeclareAbbreviation\SvAL{Soviet Astronomy Letters}
\DeclareAbbreviation\VeSon{Ver\"{o}ff. Sternw. Sonneberg}
\DeclareAbbreviation\VSOLJBul{VSOLJ Variable Star Bull.}
\DeclareAbbreviation\yCat{VizieR Online Data Catalog}
\DeclareAbbreviation\ZA{Z. Astrophys.}

\def\ASPConf#1#2{ASP Conf. Ser. #1, #2}

\def\PublisherKluwer{Dordrecht: Kluwer Academic Publishers}
\def\PublisherASP{San Francisco: ASP}
\def\PublisherReidel{Dordrecht: D. Reidel Publishing Company}

\def\PublisherUAP{Tokyo: Universal Academy Press}

\begin{document}
\SetRunningHead{T. Kato et al.}{WZ Sge-type Dwarf Nova EG Cnc}

\title{Superhumps and Repetitive Rebrightenings of the WZ Sge-Type
       Dwarf Nova, EG Cancri}

\author{Taichi \textsc{Kato}}
\affil{Department of Astronomy, Kyoto University, Sakyo-ku, Kyoto 606-8502}
\email{tkato@kusastro.kyoto-u.ac.jp}

\author{Daisaku \textsc{Nogami}}
\affil{Hida Observatory, Kyoto University,
       Kamitakara, Gifu 506-1314}
\email{nogami@kwasan.kyoto-u.ac.jp}

\author{Katsura \textsc{Matsumoto}}
\affil{Astronomical Institute, Osaka Kyoiku University, Asahigaoka,
       Kashiwara, Osaka 582-8582}
\email{katsura@cc.osaka-kyoiku.ac.jp}

\email{\rm and}

\author{Hajime \textsc{Baba}}
\affil{Center for Planning and Information Systems, The Institute of
       Space and Astronautical Science (ISAS), \\
       Sagamihara, Kanagawa 229-8510}
\email{baba@plain.isas.ac.jp}



\maketitle

\begin{abstract}
We report on time-resolved photometric observations of the WZ Sge-type
dwarf nova, EG Cnc (Huruhata's variable), during its superoutburst
in 1996--1997.  EG Cnc, after the main
superoutburst accompanied with development of superhumps typical of
a WZ Sge-type dwarf nova, exhibited a series of six major rebrightenings.
During these rebrightenings and the following long fading tail, EG Cnc
persistently showed superhumps having a period equal to the superhump period
observed during the main superoutburst.  The persistent superhumps had
a constant superhump flux with respect to the rebrightening phase.
These findings suggest the superhumps observed during the rebrightening
stage and the fading tail are a ``remnant" of usual superhumps, and are
not newly triggered by rebrightenings.  By comparison with the 1977 outburst
of this object and outbursts of other WZ Sge-type dwarf novae, we propose
an activity sequence of WZ Sge-type superoutbursts, in which the current
outburst of EG Cnc is placed between a single-rebrightening event and
distinct outbursts separated by a dip.  The post-superoutburst behavior of
WZ Sge-type dwarf novae can be understood in the presence of considerable
amount of remnant matter behind the cooling front in the outer accretion
disk, even after the main superoutburst.  We consider the premature
quenching of the hot state due to the weak tidal effect under the extreme
mass ratio of the WZ Sge-type binary is responsible for the origin of the
remnant mass.
\end{abstract}

\section{Introduction}\label{sec:intro}

   WZ Sge-type dwarf novae are an unusual subtype of SU UMa-type dwarf
novae (\cite{bai79wzsge}; \cite{dow81wzsge}; \cite{pat81wzsge};
\cite{odo91wzsge}; \cite{kat01hvvir}).  WZ Sge-type dwarf have
properties common to SU UMa-type dwarf novae
(\cite{vog80suumastars}; \cite{war85suuma}; \cite{war95suuma})
in that they show superoutbursts and superhumps.
The most unusual properties of WZ Sge-type dwarf novae are
characterized by the extremely low frequency of outbursts
(usually once in years to decades), the extremely large outburst
amplitude (exceeding 6 mag) and long (in some cases reaching $\sim$100 d
before the quiescence is reached) duration, and the lack or extreme
deficiency of normal (short) outbursts, which, in ordinary SU UMa-type
dwarf novae, occur more frequently than superoutbursts
(cf. \cite{war95suuma}).

   The origin of such unique characters has been long sought, both from
the observational and theoretical sides.  Two basic ideas have been
historically proposed, just like a descendant from historical discussions
on the mechanism of dwarf nova outbursts.  One is the mass-transfer
instability model, which assumes the outer atmosphere of the extreme
low-mass secondary in WZ Sge-type dwarf novae become unstable to the
irradiation from the outbursting accretion disk, thus giving rise to
an enhanced mass-transfer from the secondary, which leads to a long-lasting
superoutburst (the best historical example of this application being
\cite{pat81wzsge}).

   The other is an application of the disk-instability model, which,
with an assumption of very low quiescent viscosity of the accretion disk,
can produce the main characteristics of WZ Sge-type dwarf novae, namely
the long outburst intervals and the absence of normal outbursts
(\cite{sma93wzsge}; \cite{osa95wzsge}).
The required quiescent viscosity ($\alpha_C$) to reproduce such
long outburst intervals has been shown to be extremely small
($\alpha_C < 0.00005$: \cite{sma93wzsge};
$\alpha_C < 0.003$: \cite{osa95wzsge}).
These values of the required quiescent viscosity has usually been
considered too small for a disk with a developed magneto-rotational
instability \citet{BalbusHawley}, although this problem is recently
becoming more positively resolved by considering a suppression of
this instability in a cold disk with a low electric conductivity
\citet{gam98}.

   The apparent difficulty with the low quiescent viscosity has
historically invoked modifications of the disk-instability model
exemplified by the truncated inner disk \citet{las95wzsge}.
This model supposed a cold steady state-solution during quiescence,
and a slight variation in mass-transfer rate was supposed to be the
cause of infrequent outbursts.  \cite{war96wzsge} proposed another
possibility of the magnetically truncated disk.
Although these truncated disks were successful in reproducing
one of the characteristics of WZ Sge-type dwarf novae
(the long recurrence time), these models with a high $\alpha_C$
($\sim$ 0.01) have been shown by \cite{osa98suumareviewwzsge}
to meet difficulties in reproducing long-lasting superoutbursts.
Although there have been attempts (e.g. \cite{bua02suumamodel})
to reproduce long outbursts by considering an enhanced mass-transfer
during superoutbursts, the resultant light curve is very different
from observation.  Furthermore, the supposed observational evidence
for an enhanced mass-transfer is excluded by later careful analysis
\citep{osa03DNoutburst}.

   In addition to these outburst properties, WZ Sge-type dwarf novae
are shown to have other properties common to all members,
but are absent in ordinary SU UMa-type dwarf novae.  One is the presence of
{\it early superhumps}\footnote{
  This feature is also referred to as {\it orbital superhumps}
  \citep{kat96alcom} or {\it outburst orbital hump} \citep{pat98egcnc}.
} during the earliest stage of superoutbursts of WZ Sge-type dwarf novae
(cf. \cite{kat02wzsgeESH}).  The early superhumps have
a stable period almost identical with the orbital period [during the best
observed 2001 superoutburst of WZ Sge, the period of the early superhumps
was reported to be very slightly shorter than the orbital period
\citep{ish02wzsgeletter}].

   This feature was historically first detected and described in
WZ Sge itself (\cite{boh79wzsge}; \cite{pat81wzsge}), although the origin
of this variation was not properly discussed or identified at this time.
The common existence of early superhumps among WZ Sge-type dwarf novae
has been progressively confirmed:
AL Com (\cite{kat96alcom}; \cite{how96alcom}; \cite{pat96alcom};
\cite{nog97alcom}; \cite{ish02wzsgeletter}),
HV Vir (\cite{kat01hvvir}; \cite{ish03hvvir}),
RZ Leo (\cite{ish01rzleo}), and the main focus
of this paper, EG Cnc (\cite{mat96egcnciauc}; \cite{mat98egcnc}).
The other is occasional ``dips" (transient fading episodes lasting
for one to several days) during the later stage of superoutburst,
which was first clearly described in the 1978 outburst of WZ Sge
(e.g. \cite{ort80wzsge}),\footnote{
  \citet{ric92wzsgedip} was the first to discuss on the systematic tendency
  of the existence of ``dips" in outbursts for stars with infrequent
  outbursts.  This discussion, however, has been considered rather ambiguous
  mainly from the imaginary interpolation of the data, and
  from the general lack of knowledge of that time regarding individual
  dwarf-nova subtypes.
} and was repeated in the 1995 outburst of
AL Com (\cite{kat96alcom}; \cite{how96alcom}; \cite{pat96alcom};
\cite{nog97alcom}).  \citet{nog97alcom} repoeted that the stage after
the dip phenomenon shares the characteristics common to superoutbursts
of ordinary SU UMa-type dwarf novae: the presence of a likely
precursor and the subsequent growth of usual superhumps.  From these
findings, the 1977-1978 outburst of WZ Sge and the 1995 outburst of
AL Com were interpreted as ``double superoutbursts"
\citep{nog97alcom}.\footnote{
  The dwarf nova V2176 Cyg, discovered by \citet{hu97v2176cygiauc},
  probably showed the same course of double superoutbursts
  (\cite{van97v2176cygiauc}; \cite{nov01v2176cyg}).  See also
  T. Kato, vsnet-alert 1195 (1997), $\langle$http://www.kusastro.kyoto-u.ac.jp/vsnet/Mail/alert1000/msg00195.html$\rangle$.
}

   An argument naturally rises why such a sequence of superoutburst can be
only realized in WZ Sge-type dwarf novae.  A naive explanation may be a
consequence of enhanced-mass transfer, but is rather difficult to reconcile
considering the low proper angular momentum of the ``fresh" transferred
matter, accumulation of which will more easily produce normal outbursts
than a second superoutburst.  \citet{osa95wzsge}, \citet{how96alcom}, and
\citet{kuu96TOAD} instead suggested a possibility of the cooling and
heating waves originating before the termination of the entire
superoutburst.  \cite{nog97alcom} presented observational evidence in
AL Com that the second superoutburst was triggered by a short outburst,
which, from the absence of superhumps, was attributed to a normal
outburst as seen in ordinary SU UMa-type dwarf novae.
Independent observations during the second (super)outburst
are critically needed to confirm the universality of the claimed
observational features, and to explain the unique outburst pattern of
WZ Sge-type dwarf novae.

   On 1996 November 30, an alert (vsnet-alert 599)
through the VSNET alert system \citep{VSNET}
notifying a bright outburst of the suspected WZ Sge-type star,\footnote{
$\langle$http://www.kusastro.kyoto-u.ac.jp/vsnet/Mail/vsnet-alert/\\msg00599.html$\rangle$ and
$\langle$http://www.kusastro.kyoto-u.ac.jp/vsnet/DNe/egcnc.html$\rangle$.
}
EG Cnc \citep{sch96egcnciauc} announced the opening of a new story.

   EG Cnc was originally discovered by \citet{hur83egcnc}
from his photographic archive on the occasion of an outburst in 1977.
Subsequent searches for other photographic materials between
1977 and 1983, and searches for further outbursts by Huruhata himself
and a number of amateur astronomers have yielded a null result until 1995.
\citet{mcn86egcnc} correctly identified the quiescent counterpart
on Palomar Sky Survey prints, and suggested from its marked blue color
and the apparent low frequency of outbursts that this object is similar
to the dwarf nova WZ Sge.

   Upon the alert by Schmeer, we started time-resolved CCD photometry
using telescopes at Osaka Kyoiku University (OKU) and Ouda Station
\citep{Ouda}, Kyoto University.
The initial results proving the existence of
early superhumps ($P = 0.05877$ d) and the subsequent development of usual
superhumps ($P = 0.06038$ d) were already discussed in our earlier paper
(\cite{mat98egcnc}).
We discuss mainly on its peculiar late-stage phenomenon, particularly
focusing on repetitive rebrightenings.

\section{The Course of the 1996--1997 Outburst}\label{sec:outburst}

\subsection{Observations}

   The methods of observation and reduction were the same as
in \citet{mat98egcnc}.
The observations at Osaka Kyoiku University were carried out using 
the 50-cm telescope at the Cassegrain focus (focal length $= 6.0 m$) equipped
with an Astromed EEV~88200 CCD camera ($1152 \times 790$ pixels) and
a $V$-band filter.
The observations at Ouda Station were carried out using
the 60-cm telescope at the Cassegrain focus (focal length $= 4.8 m$)
equipped with a Thomson TH~7882 CCD camera ($576 \times 384$ pixels).
We used an interference filter designed to reproduce the Johnson $V$ band.
The $2 \times 2$ on-chip binning mode was used.
The frames obtained at Osaka Kyoiku University were reduced using
the IRAF.\footnote{
    IRAF is distributed by the National Optical Astronomy Observatories,
    U.S.A.
}
The frames obtained at Ouda Station were reduced with a microcomputer-based
automatic aperture and PSF photometry packages developed by one of the
authors (TK).  We used GSC 1948.1452 ($V$ = 12.82) as the local
comparison star, whose constancy during the observation was confirmed
to 0.01 mag using the check star GSC 1948.1193 ($V$ = 13.86).
Heliocentric correction were applied before the following analysis.
Table \ref{tab:log} lists the log of CCD photometry.

\begin{table*}
\caption{Journal of CCD photometry.}\label{tab:log}
\begin{center}
\begin{tabular}{ccccccc}
\hline\hline
\multicolumn{2}{c}{Date} & Start--End (UT) & $N$ & Exp(s)
        & Mean $V$ mag & Obs\commenta \\
\hline
1996 & December &  2.625 --  2.791 &  195 & 30-90  & 12.38 & OKU \\
     &          &  3.625 --  3.708 &   66 & 60-80  & 12.51 & OKU \\
     &          &  7.679 --  7.811 &  208 & 45     & 12.42 & Ouda \\
     &          &  8.715 --  8.870 &  176 & 60     & 12.48 & Ouda \\
     &          &  9.629 --  9.755 &  195 & 45     & 12.54 & Ouda \\
     &          & 11.635 -- 11.870 &  345 & 45     & 12.76 & Ouda \\
     &          & 13.746 -- 13.877 &  166 & 60     & 13.18 & Ouda \\
     &          & 14.683 -- 14.882 &  322 & 45     & 14.32 & Ouda \\
     &          & 20.666 -- 20.860 &  112 & 120    & 16.13 & Ouda \\
     &          & 21.550 -- 21.871 &  478 & 45-60  & 13.29 & Ouda \\
     &          & 22.770 -- 22.824 &   49 & 80-90  & 14.10 & Ouda \\
     &          & 23.555 -- 23.856 &  264 & 90     & 15.39 & Ouda \\
     &          & 24.555 -- 24.870 &  277 & 90     & 15.92 & Ouda \\
     &          & 25.586 -- 25.697 &  102 & 60     & 15.59 & Ouda \\
     &          & 26.700 -- 26.883 &  140 & 45-60  & 16.23 & Ouda \\
     &          & 27.619 -- 27.883 &  322 & 60     & 16.22 & Ouda \\
     &          & 30.600 -- 30.871 &  320 & 45-90  & 15.84 & Ouda \\
     &          & 31.519 -- 31.745 &  195 & 90     & 16.30 & Ouda \\
1997 & January  &  2.543 --  2.753 &  144 & 90     & 16.25 & Ouda \\
     &          &  3.862 --  3.863 &    2 & 120    & 16.35 & Ouda \\
     &          &  4.473 --  4.888 &  721 & 40-50  & 13.37 & Ouda \\
     &          &  6.616 --  6.730 &   72 & 90     & 15.98 & Ouda \\
     &          & 10.725 -- 10.795 &   82 & 60     & 13.92 & Ouda \\
     &          & 11.640 -- 11.804 &   97 & 90     & 15.29 & Ouda \\
     &          & 12.614 -- 12.755 &   78 & 90     & 16.32 & Ouda \\
     &          & 13.605 -- 13.751 &  122 & 90     & 16.23 & Ouda \\
     &          & 15.647 -- 15.811 &  137 & 90     & 16.27 & Ouda \\
     &          & 19.756 -- 19.745 &   75 & 90     & 16.35 & Ouda \\
     &          & 20.501 -- 20.843 &  284 & 90     & 16.22 & Ouda \\
     & February &  1.664 --  1.669 &    5 & 90     & 17.17 & Ouda \\
     &          &  3.433 --  3.439 &    5 & 120    & 17.38 & Ouda \\
     &          &  5.443 --  5.454 &    6 & 180    & 17.29 & Ouda \\
     &          & 27.512 -- 27.697 &   87 & 90-120 & 17.82 & Ouda \\
     &          & 28.490 -- 28.617 &   61 & 150    & 17.78 & Ouda \\
     & March    &  1.619 --  1.727 &   57 & 150    & 17.83 & Ouda \\
     &          &  3.589 --  3.726 &   63 & 150    & 17.95 & Ouda \\
     &          &  4.634 --  4.727 &   49 & 150    & 17.94 & Ouda \\
     &          &  5.522 --  5.680 &   82 & 150    & 18.02 & Ouda \\
     &          &  7.631 --  7.639 &    4 & 150    & 17.92 & Ouda \\
     &          &  8.493 --  8.675 &   96 & 150    & 18.01 & Ouda \\
     &          & 11.486 -- 11.616 &   58 & 180    & 18.09 & Ouda \\
     &          & 28.431 -- 28.516 &   39 & 150    & 17.94 & Ouda \\
     &          & 30.503 -- 30.600 &   50 & 150    & 18.25 & Ouda \\
     &          & 31.469 -- 31.604 &   61 & 180    & 18.26 & Ouda \\
     & April    &  1.435 --  1.602 &   58 & 240    & 18.15 & Ouda \\
     &          & 13.455 -- 13.463 &    5 & 150    & 17.97 & Ouda \\
     &          & 14.467 -- 14.472 &    4 & 150    & 17.73 & Ouda \\
     &          & 24.442 -- 24.450 &    5 & 170    & 18.45 & Ouda \\
     &          & 25.459 -- 25.463 &    3 & 150    & 18.55 & Ouda \\
\hline
 \multicolumn{7}{l}{\commenta OKU: Osaka Kyoiku University,
                    Ouda: Ouda Station (Kyoto University).} \\
\end{tabular}
\end{center}
\end{table*}

\subsection{Course of the 1996--1997 Outburst}

\begin{figure*}
  \begin{center}
    \FigureFile(160mm,80mm){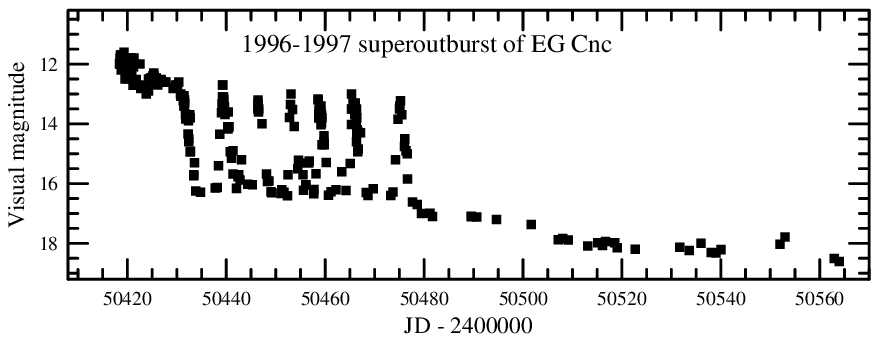}
  \end{center}
  \caption{Light curve of the 1996--1997 outburst of EG Cnc.
  Data are from visual and CCD observations reported
  to the VSNET, and $V$ measurements from this study and from
  \citet{pat98egcnc}.  The same symbols for different observers were
  chosen to best illustrate the complex light curve.
  The overall course of the outburst is characterized by the initial
  main outburst and six short outbursts (rebrightenings).
  }
  \label{fig:vis}
\end{figure*}

   Figure \ref{fig:vis} represents the overall course of the outburst from
CCD and visual observations reported to the VSNET, including the available
$V$-band CCD observations, averaged to one to a few points per night.
The object was first detected in outburst on 1996 November 30.917 UT
at $m_{\rm vis}\sim $12 \citep{sch96egcnciauc}.
The last available negative
observation was done on November 22.685 UT, giving an upper limit of
13.3 (T. Watanabe, private communication).  The early rise was
unfortunately not observed due to the interference by the Moon.
The outburst thus started some time between November 22 and 30.
The existence of early superhumps \citep{mat98egcnc}, which are
known to rapidly decay
within several days of the maximum light of a WZ Sge-type superoutburst,
and the recorded maximum brightness ($m_{\rm vis}$ = 11.8) close to that
of the 1977 outburst, suggest that the detection by Schmeer was made
within a few days of the start of the outburst.

   The main superoutburst lasted until December 15 (HJD 2450433), followed
by a precipitous decline.  The object then faded close to or
below $V$ = 16.  On December 20 (HJD 2450438), the first hint of rebrightening
was recorded in the Ouda data.  On the next night, the object reached the
peak brightness of $V$ = 13.29, and soon started fading rapidly.
This rebrightening was first considered to be of the rather common
phenomenon among short orbital-period SU UMa-type dwarf novae (e.g.
V1028 Cyg: \cite{bab00v1028cyg}, see also \cite{kat98super}),
not necessarily confined to WZ Sge-type dwarf novae.
The unique property of EG Cnc, however, emerged after this.

   The star remained about 3 mag brighter than its quiescence, and again
showed a rebrightening 7 days after its first one.  Similar rebrightenings
repetitively occurred six times in total
(counting those exceeding $V$ = 14 at maximum),
separated by five to eleven days.  Table \ref{tab:reblist} summarizes
the observed post-superoutburst rebrightenings.
Although there was a suggestion of two rebrightenings after the 1994
superoutburst of another WZ Sge-type candidate, UZ Boo
\citep{kuu96TOAD},\footnote{
  Our own examination of the available data, together with those
  presented in \citet{kuu96TOAD}, led to a less conclusive result
  regarding the separate occurrence of rebrightenings in UZ Boo.  The firm
  conclusion has been hindered by the insufficient detection threshold,
  and by the unfavorable visibility condition just before
  the solar conjunction.
}, such a large number of repetitive brightenings had been never observed
in any dwarf novae until this outburst of EG Cnc.

\begin{table}
\caption{List of rebrightenings.}\label{tab:reblist}
\begin{center}
\begin{tabular}{ccc}
\hline\hline
Date of maximum\commenta & Magnitude\commentb & $T$ (d)\commentc \\
\hline
50439 & 13.3 & $\cdots$ \\
50446 & 13.4 &  7 \\
50453 & 13.3 &  7 \\
50458 & 13.2 &  5 \\
50465 & 13.3 &  7 \\
50475 & 13.2 & 10 \\
\hline
 \multicolumn{3}{l}{\commenta JD$-$2400000.} \\
 \multicolumn{3}{l}{\commentb Maximum visual magnitude.} \\
 \multicolumn{3}{l}{\commentc Time since the preceding rebrightening.} \\
\end{tabular}
\end{center}
\end{table}

\begin{figure}
  \begin{center}
    \FigureFile(88mm,120mm){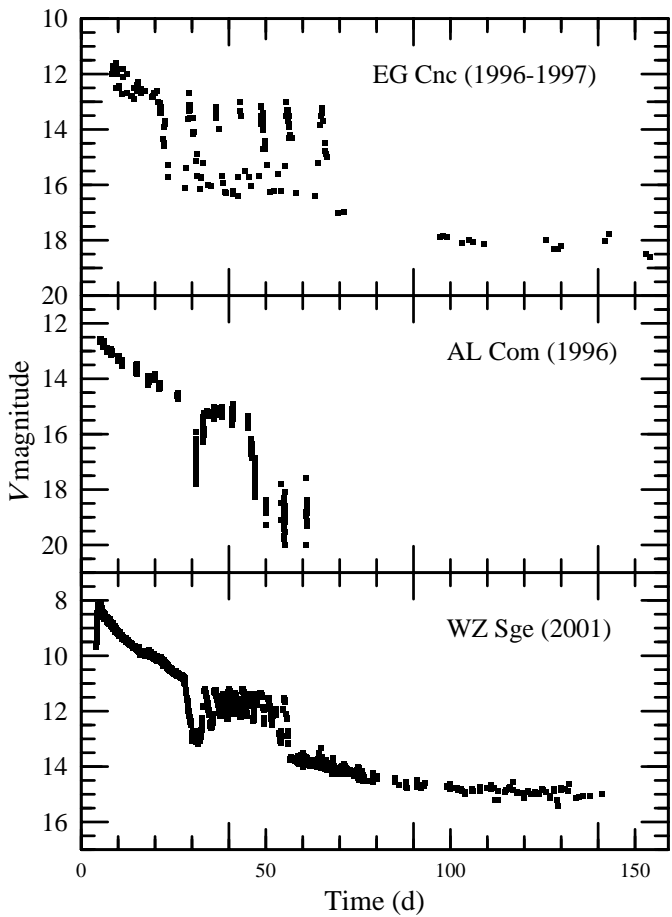}
  \end{center}
  \caption{Comparison of outburst light curves of three best observed
  WZ Sge-type dwarf novae in the recent years.
  The data of AL Com and WZ Sge are taken from \citet{nog97alcom} and
  R. Ishioka et al. in preparation, respectively.
  }
  \label{fig:wzcomp}
\end{figure}

   After its sixth rebrightening on 1997 January 26, EG Cnc then entered a
stage of slow, steady decline, keeping the rate of slow fading between
rebrightenings.  This stage lasted at least until our last observation on
1997 April 25 (146 d after Schmeer's detection
of the outburst), when EG Cnc had a magnitude of $V$ = 18.6,
which was still 0.6 mag above the quiescence (cf. \cite{mat98egcnc}).
EG Cnc thus reproduced the long-fading tail of WZ Sge observed in 1978
(cf. \cite{ort80wzsge}) and AL Com in 1995 (\cite{nog97alcom}),
accompanied by numerous short
rebrightenings instead of a superoutburst-looking second outburst
in WZ Sge and AL Com.
Figure \ref{fig:wzcomp} shows the comparison of outburst light curves
of three best-observed WZ Sge-type dwarf novae in the recent years.

\section{Superhumps and rebrightenings}

\subsection{Superhumps}

\begin{figure}
  \begin{center}
    \FigureFile(88mm,120mm){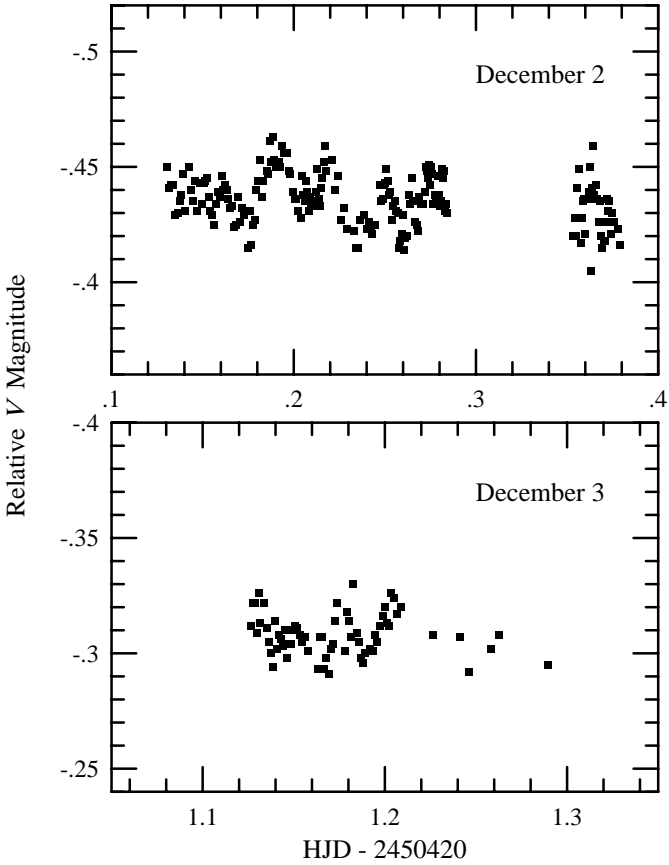}
  \end{center}
  \caption{Enlargement of the light curve of EG Cnc showing early
  superhumps.
  }
  \label{fig:earlylc}
\end{figure}

\begin{figure}
  \begin{center}
    \FigureFile(88mm,60mm){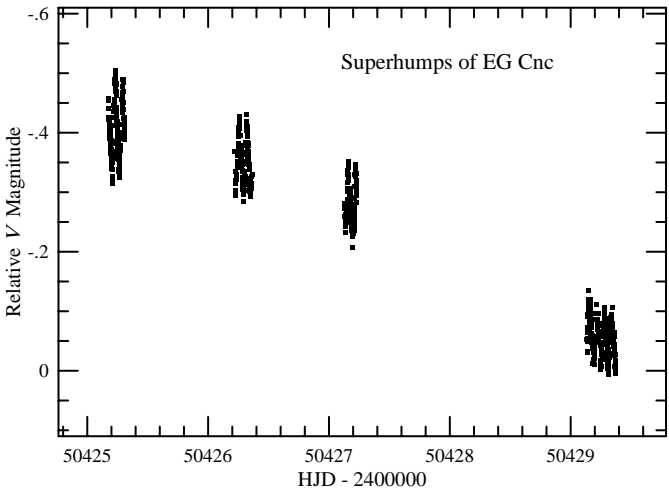}
  \end{center}
  \caption{Enlargement of the light curve of EG Cnc showing common
  superhumps.
  }
  \label{fig:shlc}
\end{figure}

   As already described in \citet{mat98egcnc}, doubly humped
early superhumps were observed during the earliest two nights of
our observations.   Since the raw light curve was not presented in
\citet{mat98egcnc}, we present the light curve in figure \ref{fig:earlylc}.
Then ``textbook" superhumps (often referred to as {\it common superhumps}
or {\it ordinary superhump} when discussing on WZ Sge-type stars)
emerged (e.g. figure \ref{fig:shlc}).

\begin{figure}
  \begin{center}
    \FigureFile(88mm,120mm){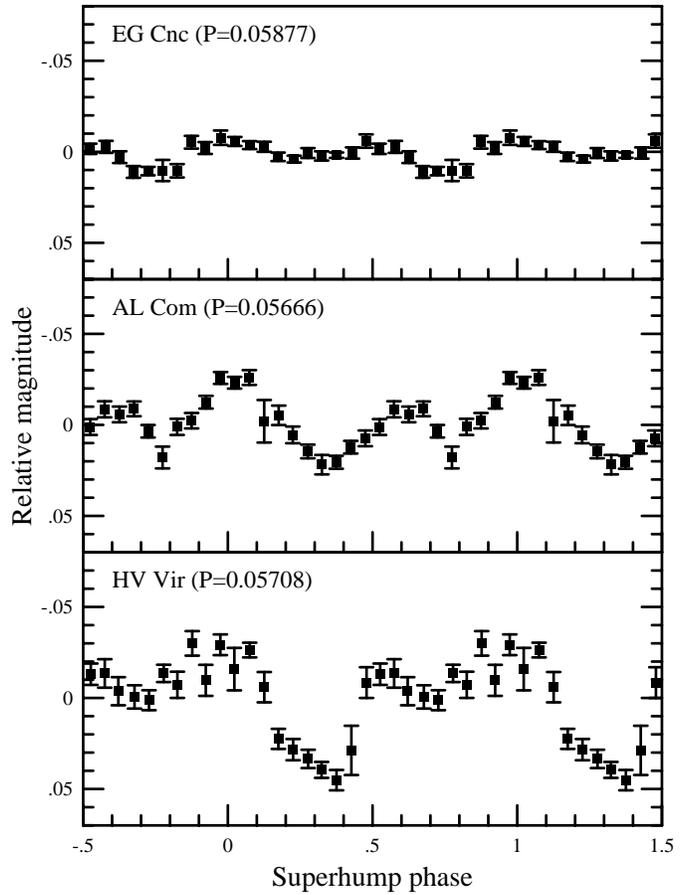}
  \end{center}
  \caption{Comparison of the profiles of early superhumps of the recent
  three well-observed WZ Sge-type dwarf novae.  Each error bar represents
  the 1-$\sigma$ error of each phase bin.
  All of the them showed doubly humped superhumps during the earliest
  stage of superoutburst.
  }
  \label{fig:earlysh}
\end{figure}

\begin{figure}
  \begin{center}
    \FigureFile(88mm,120mm){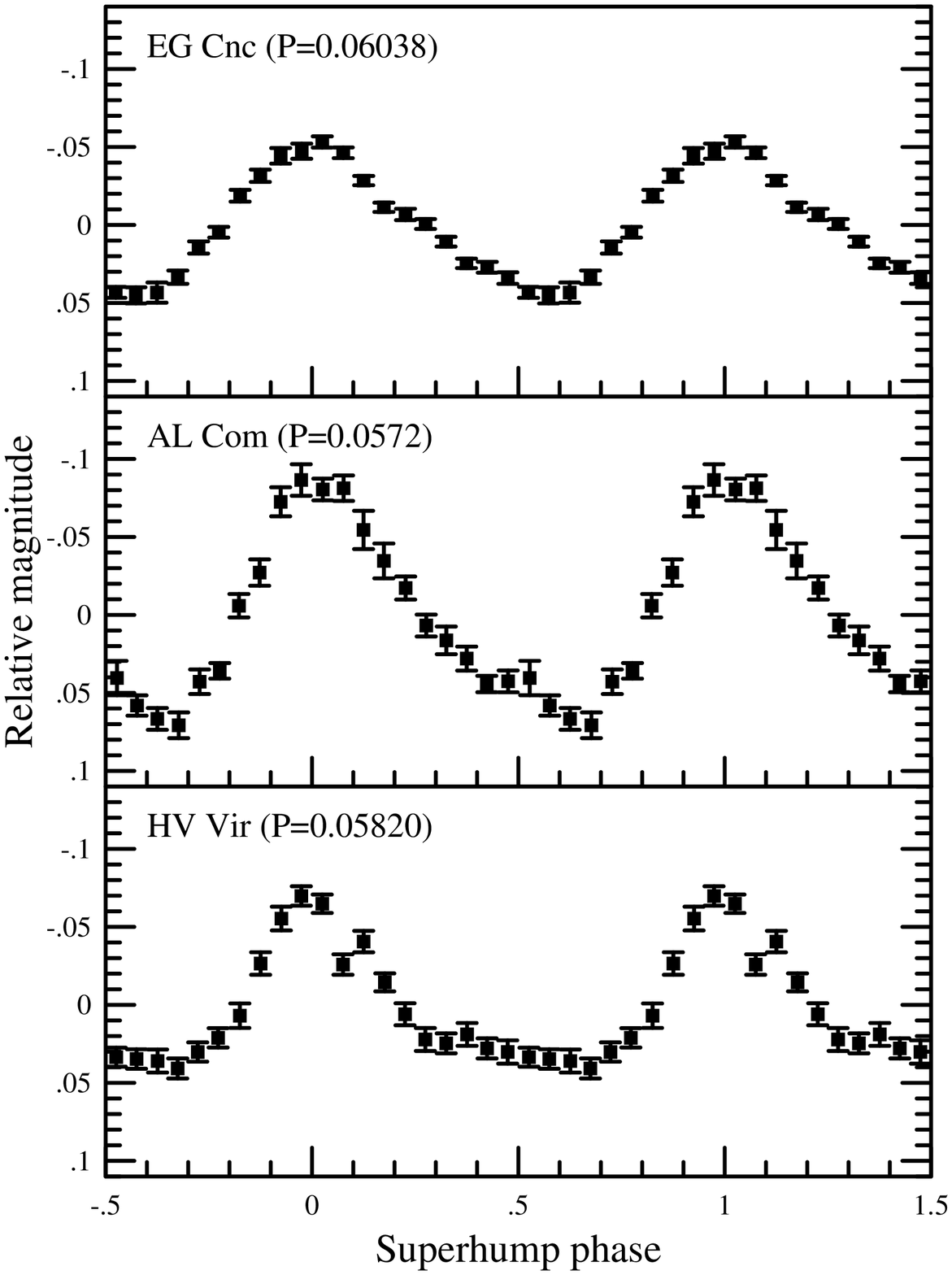}
  \end{center}
  \caption{Comparison of the profiles of common superhumps of the
  recent three well-observed WZ Sge-type dwarf novae.
  }
  \label{fig:sh}
\end{figure}

   Figures \ref{fig:earlysh} and \ref{fig:sh} show
the comparison of the profiles of these two kinds
of superhumps between the recent three well-observed WZ Sge-type dwarf
novae (the data of AL Com from \cite{kat96alcom}, HV Vir from
\cite{kat01hvvir}).  The periods (0.05877 d and 0.06038 d) used in
folding the data of EG Cnc are our revised estimates, using the PDM
(Phase Dispersion Minimization: \cite{PDM}) technique to the
datasets for 1996 December 2--3 (the stage showing  early superhumps)
and 1996 December 7--13 (common superhumps), after subtracting the trend
of linear decline in each dataset.  The new best periods were determined
using a minimum determination algorithm developed by one of the
authors (TK).  The 1-$\sigma$ errors for these periods were determined
to be 0.00017 and 0.00001 d, respectively, using Fernie's application
to Lafler-Kinman class of statistics \citep{fer89error}.
Statistical F-tests yielded significance levels of the signals 87\%
($N$ = 261) and $>$99.9\% ($N$ = 805), respectively.
As is evident from these figures, all of these three WZ Sge-type dwarf
novae showed doubly peaked early superhumps, sometimes with a rather
complex structure (cf. \cite{kat02wzsgeESH}).

   The smallness of the amplitude in EG Cnc probably suggests
that this early modulation had already started to decay at the time of
observations, or that this system has a low orbital inclination
(cf. \cite{osa02wzsgehump}).
The existence of these early superhumps, whose periods are considered
to be very close to the orbital periods
(WZ Sge: \cite{boh79wzsge}; \cite{pat81wzsge}; \cite{ish02wzsgeletter};
\cite{pat02wzsge}), AL Com: \cite{pat96alcom}),
is now regarded as an independent criterion for discriminating
WZ Sge-type dwarf novae \citep{kat02wzsgeESH}.
The common superhumps appearing later are very similar between different
objects, and have almost identical profiles to those of
ordinary SU UMa-type dwarf novae
(cf. \cite{vog80suumastars}; \cite{war85suuma}).

\subsection{Superhumps during the Rebrightening Stage and the Fading Tail}

\begin{figure}
  \begin{center}
    \FigureFile(88mm,120mm){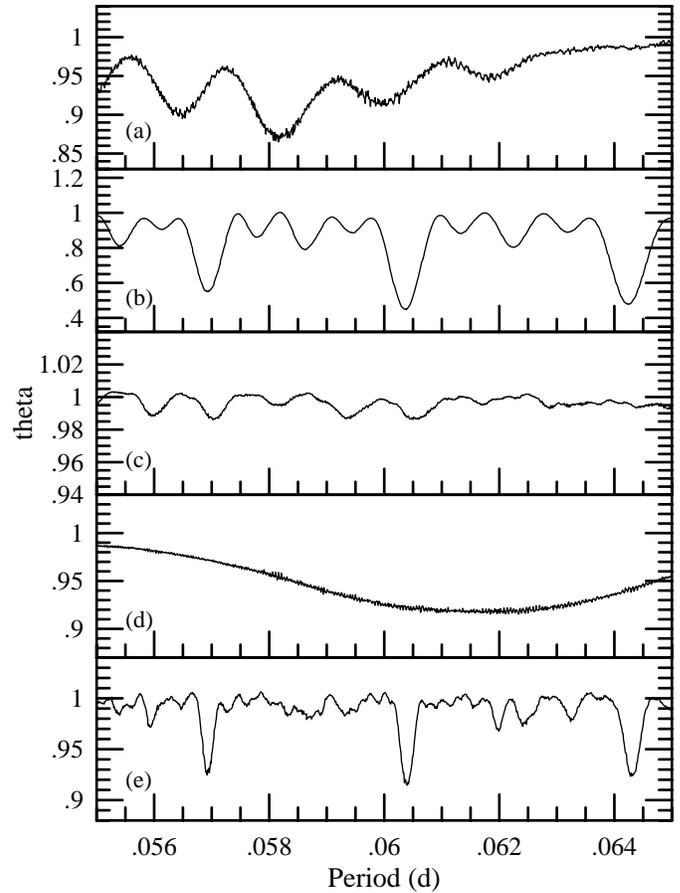}
  \end{center}
  \caption{Theta diagrams of different stages of the outburst of EG Cnc
  using PDM (Phase Dispersion Minimization, \cite{PDM}):
  (a) early superoutburst (1996 December 2 and 3),
  (b) main superoutburst (1996 December 7--13),
  (c) between rebrightenings (1996 December 26--31 and 1997 January 2),
  (d) rebrightening maximum (1997 January 4), and
  (e) fading tail (1997 February 27 and after).
  The signal (minima) around 0.0603 d represents that the
  period of ordinary superhumps of the main superoutburst.
  (The weaker signals arounds 0.0569 and 0.0643 d are one-day aliases,
  which are not true signals).
  The theta diagrams likely suggests that persistence of the 0.0603-d
  signal throughout the rebrightening phase and the fading tail of
  EG Cnc.  This can be also confirmed in the phase-averaged profiles
  shown in figure \ref{fig:fold}.
  }
  \label{fig:pdm}
\end{figure}

   We then analyzed short-term variation during the post-superoutburst
rebrightening stage.  We first divided the observations during these stages
into the following three representative stages:
(1) {\it between rebrightenings} (faint state between repetitive
rebrightenings; data for 1996 December 26-31 and 1997 January 2 were used),
(2) {\it rebrightening maximum} (around peak brightness of rebrightenings;
the 1997 January 4 data were used; note that the unique period
identification is improbable because of the short duration of the
data segment), and (3) {\it fading tail} (1997 February 27 and after).
We then removed the long-term trend of each dataset by subtracting
a linear fit, and applied the PDM technique to the datasets.
The resultant theta diagrams, together with those of the main superoutburst,
are shown in figure \ref{fig:pdm} (the weaker signals arounds 0.0569 and
0.0643 d are one-day aliases, which are not true signals).

   Although the result of the second set is less conclusive in terms of
period, the common presence of signals (minima in theta)
around 0.0603 d suggest\footnote{
  The best period determined solely from observations
  during the fading tail 0.06039(3) d, which agrees with the main superhump
  period within the estimated errors.
} that the period of usual (later) superhumps during the main superoutburst
persisted throughout the rebrightening phase and the fading tail.
Statistical F-tests yielded significance levels of the signals 59\%
($N$ = 1041), 87\% ($N$ = 716), and 84\% ($N$ = 494) for the above
three stages, respectively.

\begin{figure}
  \begin{center}
    \FigureFile(88mm,120mm){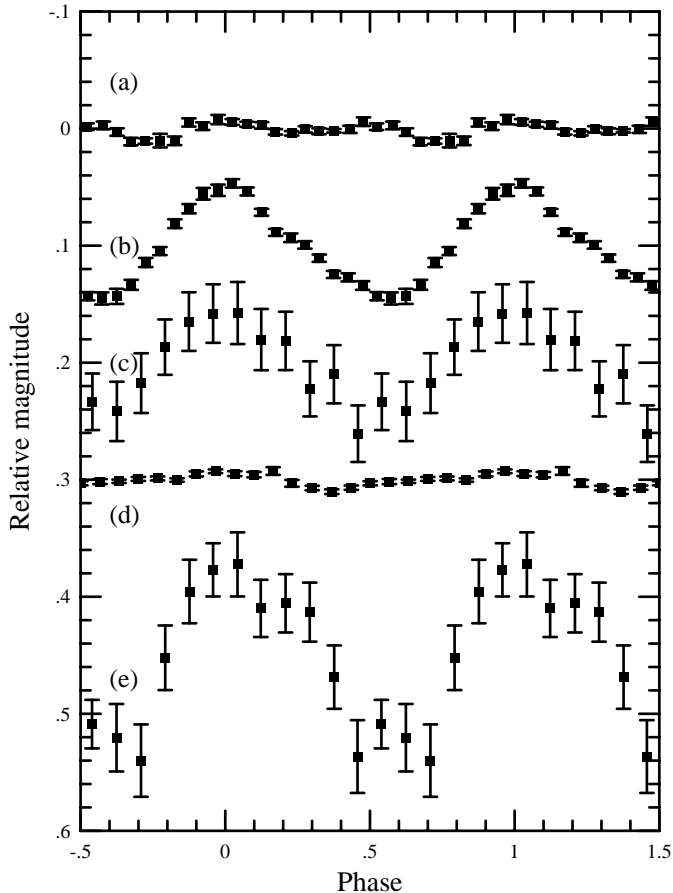}
  \end{center}
  \caption{Phase-averaged light curves of different stages of the outburst
  of EG Cnc.  The notations (a)--(e) are the same as in
  figure \ref{fig:pdm}.  The hump profiles during the rebrightening stage
  and the fading tail resembled that of usual superhumps, although
  the amplitudes were very different at different epochs.
  }
  \label{fig:fold}
\end{figure}

   Figure \ref{fig:fold} shows the phase-averaged light curves of
individual data sets.
The hump profiles, though amplitudes are very different, resemble each
other and that of usual superhumps.  This observation of strong superhump
modulation after the main superoutburst safely preclude the possibility of
the cooling white dwarf as a main source of the fading tail of WZ Sge-type
dwarf novae, as suggested by \citet{sma93wzsge}.

\begin{figure}
  \begin{center}
    \FigureFile(88mm,60mm){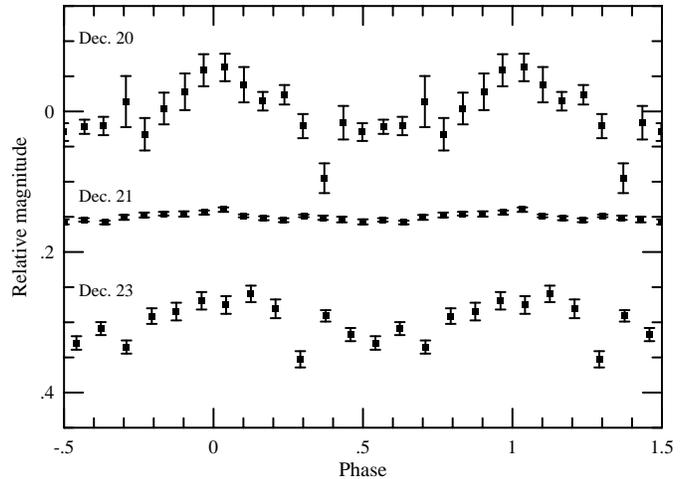}
  \end{center}
  \caption{Time-evolution of persistent superhumps with respect to the
  rebrightening stage.  The phase-averaged light curves again
  show the stability of the 0.0603-d persistent superhumps, despite that
  these persistent superhumps become weaker when the system brightens.
  }
  \label{fig:minifold}
\end{figure}

\begin{figure}
  \begin{center}
    \FigureFile(88mm,60mm){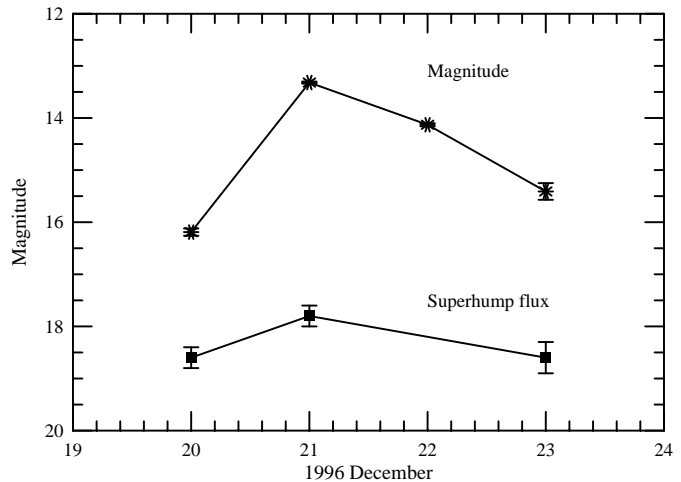}
  \end{center}
  \caption{Variation of the pulsed (superhump) flux and the total flux,
  shown in the apparent magnitude scale.
  Despite the $\sim$ 3 mag magnitude brightening of the system during
  the rebrightening, the pulsed (superhump) flux remained rather constant
  within a factor of two.
  }
  \label{fig:miniamp}
\end{figure}

   We further examined the evolution of these persistent superhumps
with respect to the rebrightening stages.
We used the data between 1996 December 20 and 23,
which best covered the rebrightening (the second rebrightening).
The resultant phase-averaged light curves are shown in
figure \ref{fig:minifold}, which again
show the stability of the 0.0603-d persistent superhumps, despite that
these persistent superhumps become weaker when the system brightens.
Figure \ref{fig:miniamp} shows the time-variation of the pulsed
(superhump) flux and the total flux of the object.  Both fluxes are
shown in magnitude scale corresponding to apparent $V$ magnitudes
(i.e. 10\% flux modulations at mean magnitude $V$ = 14.0 corresponds
to a pulsed flux corresponding to $V$ = 16.5).
Despite the $\sim$ 3 mag magnitude brightening of the system during
the rebrightening, the pulsed (superhump) flux remained rather constant
within a factor of two.

\begin{table*}
\caption{Times of superhump maxima.}\label{tab:shmax}
\begin{center}
\begin{tabular}{cccccccc}
\hline\hline
$E$\commenta & HJD$-$2450000 & Source\commentb & $O-C$\commentc
  & $E$\commenta & HJD$-$2450000 & Source\commentb & $O-C$\commentc \\
\hline
 -39   & 422.882 & 2 & 0.007  &  128   & 432.959 & 2 & -0.008 \\
 -38   & 422.943 & 2 & 0.008  &  129   & 433.020 & 2 & -0.007 \\
 -23   & 423.853 & 2 & 0.011  &  159.5 & 434.862 & 2 & -0.008 \\
  -7   & 424.809 & 2 & 0.000  &  209.5 & 437.890 & 2 & -0.002 \\
  -6   & 424.872 & 2 & 0.003  &  214.5 & 438.188 & 1 & -0.006 \\
  -5   & 424.935 & 2 & 0.005  &  215.5 & 438.252 & 1 & -0.003 \\
  -4   & 424.993 & 2 & 0.003  &  216.5 & 438.311 & 1 & -0.004 \\
  -3   & 425.055 & 2 & 0.004  &  223.5 & 438.728 & 2 & -0.010 \\
   0   & 425.236 & 1 & 0.004  &  230.5 & 439.158 & 1 & -0.003 \\
   1   & 425.299 & 1 & 0.007  &  258.5 & 440.856 & 2 & 0.003  \\
  10   & 425.839 & 2 & 0.003  &  259.5 & 440.913 & 2 & -0.000 \\
  11   & 425.900 & 2 & 0.003  &  260.5 & 440.974 & 2 & 0.000  \\
  12   & 425.961 & 2 & 0.004  &  261.5 & 441.039 & 2 & 0.005  \\
  13   & 426.022 & 2 & 0.005  &  262.5 & 441.091 & 1 & -0.004 \\
  17   & 426.263 & 1 & 0.004  &  263.5 & 441.163 & 1 & 0.008  \\
  18   & 426.324 & 1 & 0.004  &  264.5 & 441.214 & 1 & -0.002 \\
  25   & 426.743 & 2 & 0.000  &  265.5 & 441.288 & 1 & 0.012  \\
  26   & 426.805 & 2 & 0.002  &  291.5 & 442.855 & 2 & 0.008  \\
  27   & 426.861 & 2 & -0.002 &  307.5 & 443.820 & 2 & 0.006  \\
  28   & 426.923 & 2 & -0.001 &  387.5 & 448.654 & 2 & 0.005  \\
  32   & 427.165 & 1 & -0.001 &  487.5 & 454.700 & 2 & 0.008  \\
  33   & 427.225 & 1 & -0.001 &  503.5 & 455.675 & 2 & 0.017  \\
  41   & 427.707 & 2 & -0.002 &  521.5 & 456.760 & 2 & 0.014  \\
  42   & 427.768 & 2 & -0.002 &  538.5 & 457.787 & 2 & 0.014  \\
  43   & 427.828 & 2 & -0.002 &  586.5 & 460.690 & 2 & 0.016  \\
  44   & 427.892 & 2 & 0.001  &  668.5 & 465.638 & 2 & 0.009  \\
  45   & 427.950 & 2 & -0.001 &  684.5 & 466.599 & 2 & 0.003  \\
  65   & 429.162 & 1 & 0.002  &  720.5 & 468.783 & 2 & 0.011  \\
  67   & 429.282 & 1 & 0.001  &  738.5 & 469.858 & 2 & -0.002 \\
  68   & 429.343 & 1 & 0.002  &  752.5 & 470.704 & 2 & -0.002 \\
  93   & 430.846 & 2 & -0.006 &  853.5 & 476.812 & 2 & 0.003  \\
  94   & 430.909 & 2 & -0.003 &  868.5 & 477.716 & 2 & 0.001  \\
  95   & 430.970 & 2 & -0.003 &  885.5 & 478.740 & 2 & -0.003 \\
  96   & 431.031 & 2 & -0.002 &  934.5 & 481.700 & 2 & -0.004 \\
 100   & 431.272 & 1 & -0.003 & 1064.5 & 489.555 & 2 & -0.005 \\
 101   & 431.330 & 1 & -0.005 & 1083.5 & 490.699 & 2 & -0.009 \\
 104   & 431.510 & 2 & -0.007 & 1148.5 & 494.627 & 2 & -0.009 \\
 106   & 431.630 & 2 & -0.007 & 1513.5 & 516.678 & 2 & -0.015 \\
 107   & 431.691 & 2 & -0.007 & 1544.5 & 518.561 & 2 & -0.005 \\
 108   & 431.753 & 2 & -0.005 & 1613.5 & 522.717 & 2 & -0.019 \\
 110   & 431.870 & 2 & -0.009 & 1762.5 & 531.733 & 2 & -0.007 \\
 111   & 431.932 & 2 & -0.008 & 1794.5 & 533.669 & 2 & -0.005 \\
 112   & 431.992 & 2 & -0.008 & 1811.5 & 534.705 & 2 & 0.004  \\
 116   & 432.233 & 1 & -0.009 & 1845.5 & 536.755 & 2 & -0.001 \\
 117   & 432.295 & 1 & -0.007 & 1853.5 & 537.252 & 2 & 0.013  \\
 118   & 432.355 & 1 & -0.008 & 1861.5 & 537.727 & 2 & 0.004  \\
 125   & 432.780 & 2 & -0.006 & 1877.5 & 538.701 & 2 & 0.011  \\
 126   & 432.842 & 2 & -0.004 & 1886.5 & 539.240 & 2 & 0.007  \\
 127   & 432.900 & 2 & -0.006 &        &         &   &        \\
\hline
 \multicolumn{8}{l}{\commenta $E >$159 are late superhumps.} \\
 \multicolumn{8}{l}{\commentb 1. this work, 2. \citet{pat98egcnc}.} \\
 \multicolumn{8}{l}{\commentc 1. Against equation (\ref{equ:reg1}).} \\
\end{tabular}
\end{center}
\end{table*}

   We then estimated the times of superhump maxima by eye examination of
the light curve (except at the peak of rebrightening) and from the
phase-averaged light curve (around the rebrightening maximum).
The values, together with the literature values, are listed in
table \ref{tab:shmax}, with cycle counts ($E$)
starting with $E$ = 0 at HJD 2450425.236.
The accuracy of maximum determination is typically 0.001 d during the main
superoutburst, and 0.002 d during the rebrightening stage.

   In order to see whether the observed maximum times can be expressed
by a continuous period change of a coherent variation, or we need to
introduce a discontinuous phase jump (cf. \cite{kat03erumaSH}),
we calculated residuals to linear
fits to the observed times in both cases assuming the superhumps having
no phase jump and a 0.5 phase jump.  The latter (0.5 phase jump, or
phase reversal) corresponds to the phenomenon
what is called late superhumps
(\cite{hae79lateSH}; \cite{vog83lateSH}; \cite{vanderwoe88lateSH};
\cite{hes92lateSH}).

\begin{figure}
  \begin{center}
    \FigureFile(88mm,60mm){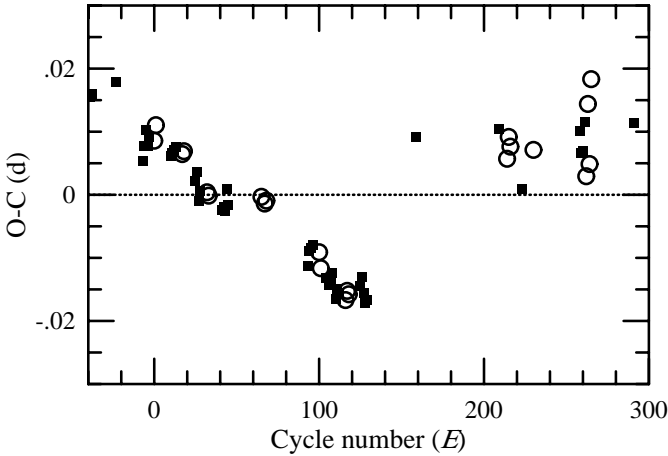}
  \end{center}
  \caption{$O-C$ diagram of the superhump maxima not assuming a phase
  0.5 jump.  The open circles
  and filled squares represent observations from this paper and from
  \citet{pat98egcnc}.  The large $O-C$ values and a discontinuous $O-C$
  change is apparent when compared to lower panel of
  figure \ref{fig:oc2}.
  }
  \label{fig:oc1}
\end{figure}

\begin{figure}
  \begin{center}
    \FigureFile(88mm,120mm){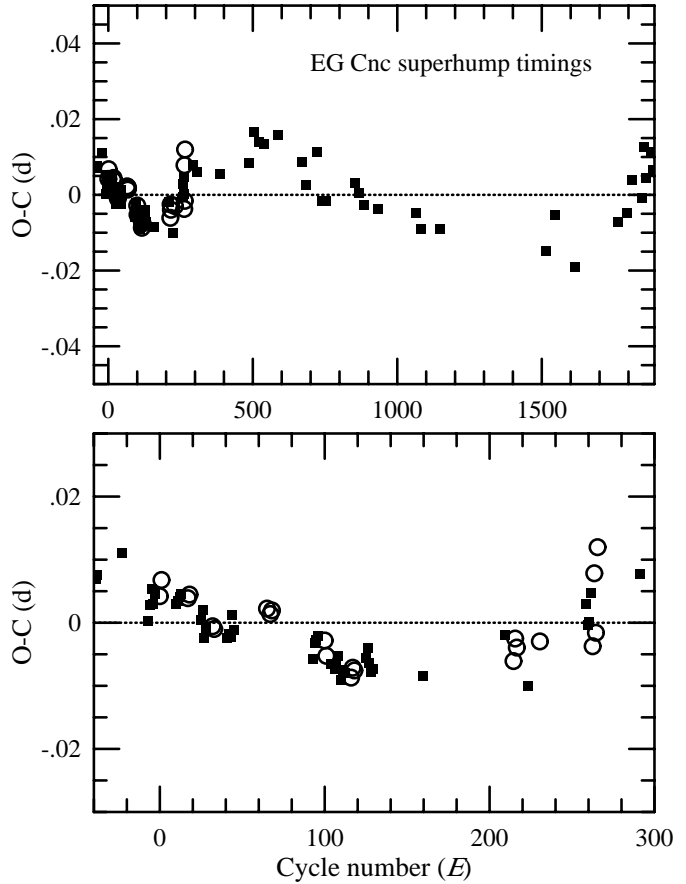}
  \end{center}
  \caption{$O-C$ diagram of the superhump maxima.  The open circles
  and filled squares represent observations from this paper and from
  \citet{pat98egcnc}.  The $O-C$'s are taken from table \ref{tab:shmax},
  assuming a phase 0.5 jump (corresponding to the appearance of late
  superhumps) between $E$ = 129 and $E$ = 159.
  (Upper:) all data.
  (Lower:) enlargement of initial stage of the main superoutburst
  and the first rebrightening.
  }
  \label{fig:oc2}
\end{figure}

   In the former case, the superhump period was confirmed to show
a discontinuous change and larger $O-C$ values (figure \ref{fig:oc1}),
while the latter case showed a more
reasonable, smooth  period change when we assume $E >$159 maxima are late
superhumps.  This identification looks reasonable that $E$ = 159
corresponds to the termination of the main superoutburst, at which
a 0.5 phase jump is known to be most frequently observed
(e.g. \cite{hae79lateSH}; \cite{vog83lateSH}).  In table
\ref{tab:shmax}, we provide $E$ numbers assuming the 0.5 phase jump
for $E >$159.

   A linear regression to the observed superhump times
(table \ref{tab:shmax}) yielded the following ephemeris:

\begin{equation}
{\rm HJD (maximum)} = 2450422.8750 + 0.060430 E. \label{equ:reg1}
\end{equation}

   The $O-C$'s given in table \ref{tab:shmax} were calculated against
this linear regression (see figure \ref{fig:oc2}).
The superhump period initially increased during the main superoutburst
and the first rebrightening,
and the period started to decrease following the main superoutburst.
A similar sequence of the superhump period change is frequently
observed in other WZ Sge-type dwarf novae or infrequently outbursting
SU UMa-type dwarf novae (e.g. \cite{kat01wxcet}; \cite{ish03hvvir};
\cite{uem03xzeri}).

   A quadratic fit to the observed $O-C$'s for $E <$210 yielded a
definitely positive period derivative of $\dot{P}/P = +1.7(1) 10^{-5}$.

   To summarize, the 0.0603-d superhump signal persisted throughout the
rebrightening stage and the fading tail.  Rebrightenings do not seem to
affect the profile, superhump phase and pulsed flux.  These findings
make a clear contrast to what are observed during the initial stages
of an ordinary SU UMa-type superoutburst \citep{war85suuma}.
The repetitive brightenings less likely represent the start of
another superoutburst in spite of accompanying superhumps.  Instead, they
bear more characteristics of normal outbursts.  The persistent superhumps
should be rather considered to be the consequence of a persistent
eccentricity in the accretion disk, and rebrightening must have occurred
independent of these superhumps.

\section{Discussion}

\subsection{Outburst Statistics}

   As briefly discussed in section \ref{sec:intro},
no confirmed outburst was observed since the 1977 (discovery) outburst
until 1995.  In order to estimate
the possibility of missed outbursts (owing to the observational gaps)
we applied Monte-Carlo simulations on actual observations by the VSOLJ
members.  The total number of observations was 238, distributing between
JD 2445344 and 2450933, spanning $\sim$15 yr.  The results were that
52\% of simulated superoutbursts (the light curve was assumed to identical
with the present one) and 86\% of simulated normal outbursts (maximum
magnitude of 13.3 and decay rate of 1 mag d$^{-1}$ assumed) were missed.
These probabilities can be considered as upper limits, since the VSOLJ
database contains no observations by Huruhata himself, and the object
had been equally intensively monitored by a number of skilled visual
observers, including P. Schmeer.  Though it is still premature to conclude
on the previous occurrence of outbursts, the 1-$\sigma$ lower limit of
recurrence times of superoutbursts ($T_{\rm s}$)
and normal outbursts ($T_{\rm n}$) can
be set 4 and 1 yr, respectively, assuming the Poisson statistic of the
outburst occurrence.  The low number of outbursts has been also confirmed
by the absence of a confirmed outburst in the VSNET observations
after the 1996--1997 one.

   These lower limits of recurrence times are located at the longest
end of the distribution of $T_{\rm s}$ among all SU UMa-type
dwarf novae (e.g. \cite{nog97sxlmi}; \cite{kat03hodel}).

\subsection{Comparison with the 1977 Outburst}

   According to \citet{hur83egcnc},
the 1977 outburst started between 1977
November 9.8 and 12.8 UT.  The brightest measured magnitude was 11.9
on 1977 November 12.8.  Since the photometric sequence of the comparison
stars by Huruhata closely matches the present scale (M. Huruhata, private
communication), we can safely compare Huruhata's values with our modern
$V$-magnitudes.  The object declined to 12.4 mag on November 22.8 UT,
10.9 d after the initial detection.  The duration of the outburst
indicates that the 1977 outburst was also a superoutburst.
After an observational gap of 11 d, \citet{hur83egcnc} further positively
detected the variable on three films taken in two separate occasions in
early December, giving estimates $\sim$14.0 mag 21--22 d and 25 d
after the initial detection.

   Considering that the duration of the main superoutburst in 1996 did
not exceed 23 d, even taking the observational gap into account,
some of these December positive detections may be interpreted as a hint of
rebrightenings, or a part of a second (super)outburst.
The main difference from the 1996--1997 outburst is
the faintness (14.0 mag) during this stage.  It is, however, clear no
evidence was found regarding repeated rebrightenings as in 1996--1997,
and the 1977 outburst seems to more resemble the 1913 and 1946 outbursts
of WZ Sge.  It would be important to note the same star at times exhibit
different patterns of activity.

\subsection{Orbital Period}

   Based on our own analysis, we propose the period of early superhumps
($P = 0.05877$ d) to be the candidate orbital period.
\citet{pat98egcnc} alternatively proposed a different (0.05997 d) period,
which is extremely close to the superhump period.
We suspect that this period identification suffered from ambiguity
from two reasons.  The spectroscopic period
by \citet{pat98egcnc} was {\it not} directly from radial velocity
variations but from a periodic variation of the asymmetry of the profile.
We must note that this method of period determination could have suffered
from the remaining eccentricity in the disk, as would be naturally
expected from the long persistence of the superhumps.
The photometric method was even more ambiguous; our own analysis of
the early-stage variations
gave a different result \citep{mat98egcnc}, which was also favored by
quiescent photometry \citet{mat98egcncqui}.  We absolutely need
an independent determination from {\it true} radial velocity observation
before concluding on the true identification of the orbital period.
In this paper, we adopt our own identification ($P = 0.05877$ d)
in the following discussion.

\subsection{EG Cnc as a WZ Sge-Type Dwarf Nova}

\begin{table*}
\caption{Comparison of WZ Sge-type dwarf novae.}\label{tab:wzcomp}
\begin{center}
\begin{tabular}{lcccc}
\hline\hline
                  & EG Cnc\commenta & AL Com\commentb
                    & HV Vir\commentc & WZ Sge\commentd \\
\hline
Year of outbursts & 1977, 1996 & 1892, 1941, 1961  & 1929, 1939, 1970, &
                    1913, 1946, 1978, \\
                  &            & 1965, 1974, 1975, & 1981, 1992        & 2001 \\
                  &            & 1995, 2001        &                   & \\
Maximum magnitude & 11.9 & 12.0 & 11.5 &  7.0 \\
Minimum magnitude & 19.1 & 20.5 & 19.1 & 15.5 \\
Outburst amplitude (mag) & 7.2 & 8.5 & 7.6 & 8.5 \\
Supercycle (yr)   & 19?  & 10--20 & 10: & 23--33 \\
Normal outbursts? & no?  & yes  & yes? & no \\
Superoutburst duration (d) & $>$100 & $>$70 & $>$50 & 130 \\
Long fading tail       & yes  & probably yes & yes & yes \\
Type(s) of rebrightening & repeated short & superoutburst-like &
                           probably yes & long (similar \\
                     & outbursts &        & (short?) & to AL Com) \\
Early superhumps     & yes & yes & yes & yes \\
Period of early superhumps (d) & 0.05877 & 0.05666 & 0.05708 & 0.05667 \\
Period of common superhumps (d) & 0.06038 & 0.05722 & 0.05820 & 0.05721 \\
Superhump excess (\%)\commente & 2.7     & 1.0     & 2.0    & 1.0   \\
Quiescent humps   & ? & double  & ?       & double+eclipses \\
\hline
 \multicolumn{5}{l}{\commenta this paper.} \\
 \multicolumn{5}{l}{\commentb \citet{kat96alcom}; \citet{pat96alcom};
                              \citet{nog97alcom}.} \\
 \multicolumn{5}{l}{\commentc \citet{kat01hvvir}; \citet{ish03hvvir};
                              \citet{lei94hvvir}.} \\
 \multicolumn{5}{l}{\commentd \citet{ish02wzsgeletter}; \citet{pat02wzsge};
                              R. Ishioka et al. in preparation.} \\
 \multicolumn{5}{l}{\commentd Calculated from the periods of early
                    superhumps and common superhumps except for WZ Sge.}
\end{tabular}
\end{center}
\end{table*}

   As \citet{mat98egcnc} pointed out, the optical behavior of the
present outburst bears all the characteristics common to those of known
WZ Sge-type dwarf novae.  Table \ref{tab:wzcomp}
summarizes the comparison of the
observed properties between these objects.  The late-stage behavior of
superoutbursts has a wide range of diversity between objects, and even
between different outbursts of the same object, in spite of the marked
similarity of the general light curve and the time evolution of superhumps
during the early stage of outbursts.

   Phenomenologically, the superoutbursts of WZ Sge-type dwarf
novae seem to make a continuous sequence ranging from superoutburst
without noticeable rebrightenings (WZ Sge in 1913 and 1946), ones
accompanied by a single rebrightening (EG Cnc in 1977; possibly HV Vir
in 1992, cf. \cite{kat01hvvir}), double to multiple rebrightenings
(EG Cnc in 1996--1997; UZ Boo in 1994: \cite{kuu96TOAD}), and to
two distinct outbursts separated by a ``dip"
(WZ Sge in 1978--1979, 2001; AL Com in 1961, 1995 and 2001),
some of which may be double superoutbursts.

   A similar idea of a continuum of types of outbursts of large-amplitude,
infrequently outbursting dwarf novae (Tremendous Outburst Amplitude
Dwaef Novae or TOADs: \cite{how95TOAD})
was discussed in \citet{how95swumabcumatvcrv},
who mainly treated the diversity
of duration and brightness of superoutbursts in these systems.
An ideal model of WZ Sge-type dwarf novae should reproduce this wide
and continuous variety of outburst activity by changing some input
parameters.

\subsection{Are rebrightenings {\it inside-out}-type or
           {\it outside-in}-type outbursts?}

   In observationally interpreting the rebrightening phenomenon, discrimination
of {\it inside-out}-type (heating wave originating from the inner
accretion disk), or {\it outside-in}-type outbursts
(\cite{can86DNburst}; the terminology corresponding to type B and type A
outbursts in \cite{sma84DI}),
is potentially important: different parameters of
the disk instability model predict different origins of the instability
(e.g. \cite{osa96review}; \cite{can96DN}),
which may provide a potential observational
diagnostic to the underlying mechanisms.

\begin{figure}
  \begin{center}
    \FigureFile(88mm,120mm){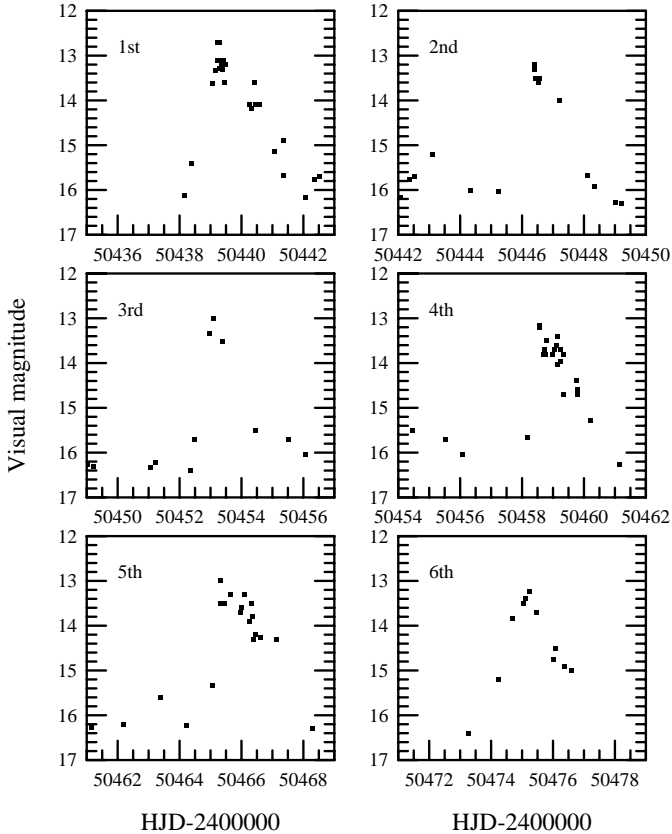}
  \end{center}
  \caption{Enlarged light curves of each rebrightening.
  The data are from VSNET visual observations and $V$-band observations
  from this work.
  The first five rebrightenings seem to show rather
  common outburst behavior: an abrupt rise and a 3-mag fade in three days.
  The time needed for the rise is well confined to less than 0.4 day
  from the 4th and 5th rebrightenings.
  The sixth rebrightening seems to be different, showing a slow rising
  trend more than a day before reaching maximum.
  }
  \label{fig:minicomp}
\end{figure}

   A close examination of the light curve of each rebrightening
revealed the presence of five rebrightenings which showed a rapid rise
and one which showed a much slower rise.
Figure \ref{fig:minicomp} demonstrates enlarged light curves of
each rebrightening.
The first five rebrightenings showed rather common outburst
behavior: an abrupt rise and a 3-mag fade in 3 d.
The time needed for the rise is well confined to less than 0.4 d
from the observations of the fourth and fifth rebrightenings.
The sixth rebrightening is different, accompanied with a slow
rise lasting more than a day before reaching maximum.

\begin{figure}
  \begin{center}
    \FigureFile(88mm,60mm){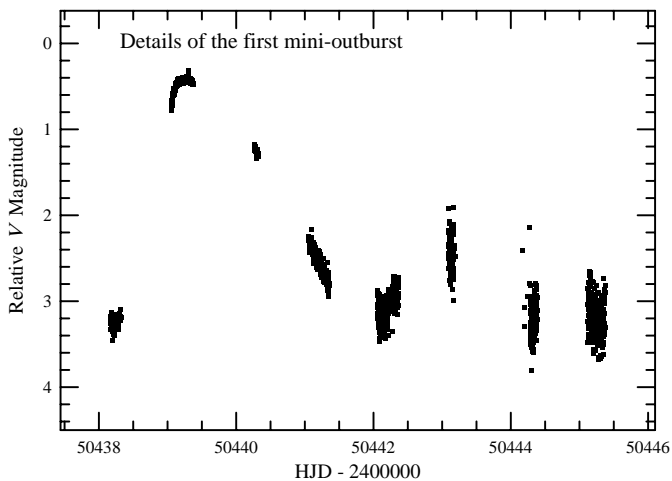}
  \end{center}
  \caption{Light curve showing the first rebrightening (HJD 2450439)
  and an immediately following small rebrightening (HJD 2450443).
  }
  \label{fig:reb1}
\end{figure}

\begin{figure}
  \begin{center}
    \FigureFile(88mm,60mm){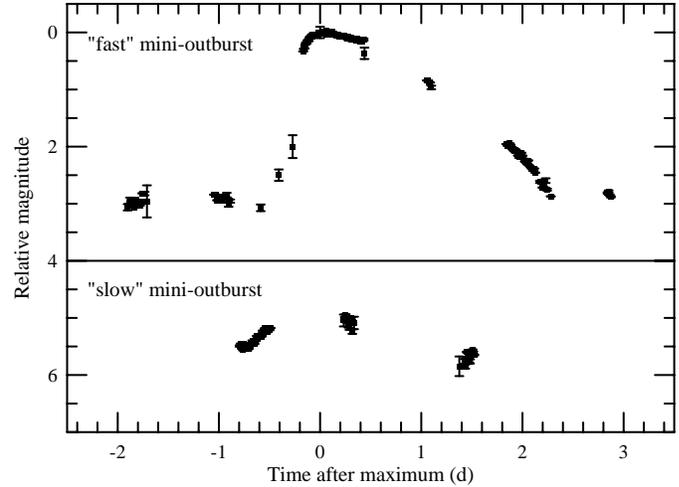}
  \end{center}
  \caption{Enlarged light curve of the first rebrightening and a small
  rebrightening immediately following the first rebrightening.
  The rebrightening occurring around HJD 2450439 represents a ``fast"
  rebrightening, while the small one (not counted as one of six major
  rebrightenings because of the faintness) occurring around HJD 2450443
  shows a much slower rise and fainter peak brightness.
  }
  \label{fig:miniprof}
\end{figure}

   A similar phenomenon, but with a lesser amplitude, was also observed
in time-resolved CCD photometry of the first rebrightening
(figures \ref{fig:reb1} and \ref{fig:miniprof}).
A fast-rising type rebrightening occurred at around HJD 2450439,
while a more slowly rising brightening
followed at around HJD 2450443 (not counted as one of six major
rebrightenings because of its faintness) shows much a slower rise and
faint peak brightness (the faintness of the peak was also confirmed from
the subsequent CCD observations reported to the VSNET).\footnote{
  We regard this phenomenon as a kind of rebrightening, not a result
  of flickering, since the duration ($\sim$ 1 d) of the phenomenon,
  as well as the continued slow rising for 7 hr, was much longer than
  the superhump period.
}  Although whether or not this small brightening can be treated in
the same context of other (major) rebrightenings is an open question,
the slow rising stage of this small brightening was more analogous to
the sixth rebrightening, possibly implying the common underlying mechanism.

   The existence of two types (fast-rising and slow-rising)
of rebrightenings may represent the coexistence
of {\it inside-out}-type and {\it outside-in}-type outbursts, more usual
ones being the faster {\it outside-in}-type.  The sixth rebrightening may
be related to the former, ``slow" rebrightening.

\subsection{On the Interpretation of Rebrightenings}

   Since the disk instability theory predicts the cycle length of normal
outbursts, for a certain range of parameters, is roughly inversely
proportional to the mass-transfer rate (e.g. \cite{ich94cycle}),
it would be natural to attribute the increased outburst activity after
the main superoutburst to an enhanced mass-transfer somehow caused by
the superoutburst.  Since the measured short outburst intervals (5--10 d)
are one of the shortest even among frequently outbursting SU UMa-type
dwarf novae (cf. \cite{war95suuma}; \cite{nog97sxlmi}),
this interpretation would require
(assuming no change in other parameters before and after the superoutburst)
a mass-transfer rate something between ordinary SU UMa-type dwarf novae
and ER UMa stars (\cite{kat95eruma}; \cite{rob95eruma}; \cite{mis95PGCV};
\cite{nog95rzlmi}; \cite{kat99erumareview}).  The ER UMa stars are
peculiar SU UMa-type stars which are proposed to have mass-transfer rates
several times higher than in ordinary SU UMa-type dwarf novae
(\cite{osa95eruma}; \cite{osa95rzlmi}).

   Although the initial pattern and amplitudes of rebrightenings
of EG Cnc closely resembled those of ER UMa stars, the enhanced
mass-transfer and the enhanced hot spot brightness would not solely by
themselves explain the later behavior, since the brightness of the system
was shown to decrease gradually (see figure \ref{fig:vis}),
even after the sixth rebrightening.
This smooth decrease in brightness should require
{\it smoothly decreasing} frequency of rebrightenings, which is contrary to
the observation if the system brightness actually reflects the brightness
of the hot spot or the mass-transfer rate.

   \citet{osa97egcnc} instead proposed a model to reproduce these
rebrightenings, by assuming increased quiescent viscosity after the main
superoutburst.  This model, by adjusting the time dependence of the
quiescent viscosity and its radial dependence through the disk, could
at least reproduce the observed rebrightenings without an assumption
of an enhanced mass-transfer.

   \citet{osa97egcnc} originally suggested that such increased quiescent
viscosity is somehow related to the turbulent motion in the disk
resulting from the tidal instability.  They also proposed the extremely
low mass-transfer rate in WZ Sge-type dwarf novae would be responsible
for the slow circularization of the eccentric disk, leading to the slow
decrease in the quiescent viscosity after the outburst and multiple
rebrightenings never seen in usual dwarf novae.  The underlying physical
mechanism for the decay of the disk viscosity has been recently proposed
by \citet{osa01egcnc} by considering a decay of MHD turbulence under the
condition of the low magnetic Reynolds numbers in the cold
accretion disk \citep{gam98}.

   A potential difficulty of this model \citep{osa01egcnc} is that the
disk after the main superoutburst would not expand enough to trigger
a second superoutburst as reported in AL Com \citep{nog97alcom}.
[Note that the outburst of AL Com after the ``dip" was not composed of
recurring small rebrightenings, as observed in the 2001 superoutburst
of WZ Sge.]

\citet{how96alcom} and \citet{kuu96TOAD}\footnote{
  \citet{kuu96TOAD} also discussed outburst properties of
  soft X-ray transients (SXTs).  At least one of SXTs (GRO~J0422+32 = V518
  Per) definitely shows similar rebrightenings (mini-outbursts).
  The cause of such rebrightenings have been discussed
  (e.g. \cite{che93BHXNsecondarymaxima}; \cite{aug93SXTecho}).
  These authors favored the irradiation-induced
  mass-transfer in SXTs as the cause of rebrightenings.  As already
  discussed \citet{kuu96TOAD}, this effect is expected to
  be much smaller, however, in dwarf novae than in SXTs.
  See also \citet{kuu98v616mon} and \citet{kuu00wzsgeSXT} for recent
  discussions.
} proposed the idea that some material can be piled up just behind
the cooling wave, causing it to be reflected as a heating wave,
which they considered to cause an additional brightening.

   Regarding the reason why such a condition is predominantly met in
WZ Sge-type dwarf novae, or TOADs, \citet{kuu96TOAD} suggested that
the stored disk material survives being depleted by accretion, owing
to the low occurrence of normal outbursts in these systems.  However, the
degree of depletion of the disk mass during one supercycle is rather
expected to affect the initial condition of superoutburst, but not
necessarily that of the late-stage condition of superoutburst.
As \citet{osa95wzsge} has shown, the late stage of WZ Sge-type
superoutbursts is expected to follow the same time-evolution as those of
usual dwarf novae.
The initial excessive matter is effectively accreted on a viscous
time-scale; it is not still convincingly clear how the low occurrence
of normal outbursts (or its precondition) naturally explains the unique
late-stage phenomenon in WZ Sge-type superoutbursts.
Though the discussion was more limited to SXTs,
\citet{can95BXHNDI} could reproduce a similar pattern of rebrightenings
by modifying the radial dependence of the $\alpha$ viscosity.
This model, however, would suffer,
as in the model proposed by \citet{osa97egcnc}, from the depletion of
high-angular momentum matter necessary to invoke an immediate second
superoutburst.

   We alternately consider a possibility that the matter in the outer
accretion disk may be left from being accreted during the main
superoutburst, by way of the premature quenching of the hot state
in the outer disk.  We propose the weaker turbulence and resultant
heating caused by the weaker tidal instability in the outer disk than
in ordinary SU UMa-type dwarf novae can be responsible for this possibility.
This weaker tidal torque can be reasonably achieved under conditions of
extreme mass ratios in WZ Sge-type dwarf novae, which can hold
the 3:1 resonance radius well inside the Roche lobe of the
primary, or the maximum dimension of the expanded disk.
This possibility was originally proposed in 1997 by \cite{kat97egcnc}
and \citet{kat98super}, which was further explored by \citet{hel01eruma}
and \citet{osa01egcnc}.  The unusual presence of Na D absorption line
\citep{pat98egcnc} could be attributed to the cool reservoir in the
disk required by this interpretation.

   Superoutbursts in ordinary SU UMa-type dwarf novae are believed to be
triggered when the disk radius, upon the ignition of the outburst, first
reaches the radius of the 3:1 resonance during the normally outbursting
cycle \citep{osa89suuma}.
The eccentricity wave \citep{lub92SH} caused by the tidal instability
should naturally start in the outermost region of the disk, and should
propagate inward, accounting for the general decrease in the superhump
period during superoutbursts of ordinary SU UMa-type dwarf novae
\citep{lub92SH}.

   In WZ Sge-type dwarf novae, however, the absence of normal outbursts
may lead to an accumulation of a large amount of matter, leading to
more violent expansion of the disk during superoutburst enabling the
disk expanding beyond the 3:1 resonance.  The above picture then predicts
the eccentricity wave can propagate outward.  The presence
of positive period derivatives of superhump periods in some WZ Sge-type
dwarf novae and infrequently outbursting, large-amplitude SU UMa-type
dwarf novae (e.g. AL Com: \cite{nog97alcom}; V1028 Cyg: \cite{bab00v1028cyg};
SW UMa: \cite{sem97swuma}; \cite{nog98swuma}; HV Vir: \cite{ish03hvvir};
see also \cite{kat03v877arakktelpucma} for a recent summary of
period derivatives in SU UMa-type dwarf novae) would strengthen
the idea of the relative difference in the location of generation and
propagation of tidal instability between WZ Sge-type and ordinary
SU UMa-type dwarf novae.  The extremely low mass-ratios in WZ Sge-type
dwarf novae could play the key role in generating tidal instability
well inside the entire disk radius.

\section{Summary}

   We obtained extensive CCD photometry during the 1996--1997 superoutburst
of EG Cnc.  This outburst showed an unprecedented series of six major
rebrightenings spaced by 5--10 d.  During this rebrightening stage
and the following slow fading tail, photometric modulations with a period
of 0.0603 d persisted, which is equal to the superhump period observed
in the latter course of the main superoutburst.  We suggest that the
superhumps observed during the rebrightening stage and the fading tail
are a ``remnant" of usual superhumps, and are not a result
of rebrightenings.  By comparison with the 1977 outburst of this object
and outbursts of other WZ Sge-type dwarf novae, we propose an activity
sequence of WZ Sge-type superoutbursts, in which the current outburst of
EG Cnc is placed between a single-rebrightening event and
distinct outbursts separated by a dip.
We propose a new potential explanation
of post-superoutburst behavior of the WZ Sge-type dwarf novae, by
considering the effect of extreme mass ratios in these binaries on
the generation and propagation of the tidal instability.

\vskip 3mm

   We are grateful to many amateur observers for supplying their vital visual
and CCD estimates via the VSNET alert network,
and especially to Patrick Schmeer for his
detection and early notification of the long-awaited outburst.  We are
grateful to the VSOLJ staffs and members providing earlier observations.
We are also grateful to Jonathan Kemp for providing up-to-date information,
and people involved in extensive on-line discussions in {\it vsnet-chat};
part of this log can be electronically accessed at
http://www.kusastro.kyoto-u.ac.jp/vsnet/DNe/egcnc.html and its related links.
We express our sincere gratitude to Yoji Osaki and Shin Mineshige for
valuable discussions made at the Annual Meeting of the Astronomical Society
of Japan.
Part of this work was supported by a Research Fellowship of the Japan
Society for the Promotion of Science for Young Scientists
(DN).

\end{document}